\documentclass{article}
\usepackage[psamsfonts]{amssymb}
\usepackage{amsmath}
\usepackage{cite}
\usepackage{bm}
\usepackage{graphicx}
\usepackage{yhmath}
\usepackage{bbm}

\def\d{\mathrm{d}}
\def\id{\mathbf{1}}
\def\ir{\mathrm{i}}

\begin{document}
\begin{titlepage}
\begin{center}
{\large \textbf{Central force problem in space with \\ SU(2) Poisson structure}}

\vspace{2\baselineskip}

{ Taraneh Andalib\footnote{e-mail: Taraneh.Andalib@gmail.com}
~~and~~
Amir~H.~Fatollahi\footnote{e-mail: fath@alzahra.ac.ir}

}

\vspace{1\baselineskip}
{\it Department of Physics, Alzahra University, Tehran 19938-93973, Iran}

\end{center}

\vspace{1\baselineskip}

\begin{abstract}
\noindent The central force problem is considered in a three
dimensional space in which the Poisson bracket among the
spatial coordinates is the one by the SU(2) Lie algebra.
It is shown that among attractive power-law potentials it is
only the Kepler one that all of its bound-states make closed
orbits. The analytic solution of the path equation under the Kepler
potential is presented. It is shown that except the Kepler potential,
in contrast to ordinary space, all of the potentials for which \emph{all} of the
\emph{almost} circular orbits are closed are non-power-law ones.
For the non-power-law potentials examples of the numerical solutions
of the path equations are presented.
\end{abstract}

\vspace{2\baselineskip}



\end{titlepage}

\section{Introduction}
During recent years much attention has been paid to the
formulation and study the physics living on noncommutative spaces.
The motivation, among others \cite{doplicher,madore1},
partly is the natural appearance of noncommutative spaces
in some areas of physics, for example in the string
theory. In particular it has been understood that the canonical relation
\begin{equation}\label{1}
[\hat x_a,\hat x_b]=\ir\,\theta_{a\, b}\,\id,
\end{equation}
with $\theta$ is an antisymmetric constant tensor,
describes the longitudinal directions of D-branes in the presence
of a constant B-field background
as seen by the ends of open strings \cite{9908142,99-2,99-3,99-4}.
The theoretical and phenomenological
implications of possible noncommutative coordinates have been
extensively studied \cite{reviewnc}.

One direction to extend studies on noncommutative spaces is to
consider spaces where the commutators of the coordinates are not
constants. Examples of this kind are the noncommutative cylinder
and the $q$-deformed plane \cite{chai}, the so-called
$\kappa$-Poincar\'{e} algebra \cite{majid,ruegg,amelino,kappa},
and linear noncommutativity of the Lie algebra type
\cite{wess,sasak}. In the latter the dimensionless spatial
position operators satisfy the commutation relations of a Lie
algebra:
\begin{equation}\label{2}
[\hat{x}_a,\hat{x}_b]= f^c{}_{a\, b}\,\hat{x}_c,
\end{equation}
where $f^c{}_{a\,b}$'s are structure constants of a Lie algebra.
One example of this kind is the algebra SO(3), or SU(2). A special
case of this is the so called fuzzy sphere \cite{madore,presnaj},
where an irreducible representation of the position operators is
used which makes the Casimir of the algebra,
$(\hat{x}_1)^2+(\hat{x}_2)^2+(\hat{x}_3)^2$, a multiple of the
identity operator (a constant, hence the name sphere). One can
consider the square root of this Casimir as the radius of the
fuzzy sphere. This is, however, a noncommutative version of a
two-dimensional space (sphere). In \cite{batista}, a four
dimensional differential calculus has been developed corresponding
to the algebra SU(2), based on which the equations of motion
corresponding to spin 0, 1/2, and 1 particles living in that four
dimensional space have been investigated.

In \cite{0612013,fakE1,fakE2} a model was introduced in which the
representation was not restricted to an irreducible one, instead the
whole group was employed; see also \cite{jablam}. In particular the
regular representation of the group was considered, which contains all
representations. As a consequence in such models one is dealing
with the whole space, rather than a sub-space like the case of
fuzzy sphere as a 2-dimensional surface. In \cite{0612013,fakE1} basic
ingredients for calculus on a linear fuzzy space, as well as the basic
notions for a field theory on such a space, were introduced.
Models based on the regular representation of SU(2)
were treated in more detail, giving explicit forms of the tools
and notions introduced in their general forms
\cite{0612013,fakE1}. In \cite{fakE1, fakE2, spinor} the tree
and loop diagrams for self-interacting scalar and spinor fields discussed.
It is observed that models based on
Lie algebra type noncommutativity enjoy three features:
\begin{itemize}
\item They are free from any ultraviolet divergences if the group
is compact.
\item The momentum conservation is modified, in the
sense that the vector addition is replaced by some  non-Abelian
operation \cite{pal,0612013}.
\item In the transition amplitudes only the so-called planar
graphs contribute.
\end{itemize}
In \cite{fsk} the quantum mechanics
on a space with SU(2) fuzziness was examined. In particular,
the commutation relations of the position and momentum
operators corresponding to spaces with Lie-algebra
noncommutativity in the configuration space, as well as
the eigen-value problem for the SU(2)-invariant systems were studied.
The consequences of the Lie type noncommutativity of space on thermodynamical
properties have been explored in \cite{shin,fsmjmp}.

The classical motion on noncommutative space 
has attracted interests as well \cite{miao,silva}.
In particular, the central force problems on space-times
with canonical and linear noncommutativity and their
observational consequences have been the subject of different 
research works \cite{levia,zhang,kapoor,dennis,romero,mirza}.
In \cite{kfs} the classical mechanics defined on a space with
SU(2) fuzziness was studied. In particular, the Poisson structure
induced by noncommutativity of SU(2) type was investigated, for
either the Cartesian or Euler parameterization of SU(2) group. The
consequences of SU(2)-symmetry in such spaces on integrability,
were also studied in \cite{kfs}.

The purpose of the present work is to examine in more detail
the classical central force problem in a space with linear SU(2) fuzziness.
In particular, using the path equation, the conditions are obtained under
which the slightly perturbed circular orbits are stable or closed.
The differential equation is obtained that the potentials should satisfy
to guarantee that all of the perturbed circular orbits of them would be closed.
The numerical solutions of the mentioned differential equation as
well as the path equation under the obtained potentials are presented.
It is shown that among attractive power-law potentials it is
only the Kepler one that all of its bound-states make closed
orbits. The analytic solution of the path equation under the Kepler
potential is given.

The scheme of the rest of this paper is the following. In section
2, a short review of the construction in \cite{kfs} is presented.
In section 3 the circular orbits and various aspects of their perturbations are considered.
In section 4 the Kepler problem and its analytic solution is presented.
Appendix A is devoted to the derivation of the differential equation by which
the potentials with closed perturbed circular orbits are obtained.

\section{Dynamics}
The classical dynamics on a space whose Poisson structure is
originated from a Lie algebra is given in \cite{kfs}. To make this presentation
self contained, below a short review of the construction is presented.

\subsection{Basic notions}
Denote the members of a basis for the left-invariant vector fields
corresponding to the group $G$ by $\hat{x}_a$'s. These fields satisfy
the commutation relation (\ref{2}), with the structure constants
of the Lie algebra corresponding to $G$. The members of this basis
would contain the quantum mechanical (operator form) counterparts
of the classical spatial coordinates of the system (for the moment they are dimensionless).
The group elements of $G$ are parametrized by the coordinates $k^a$'s as:
\begin{equation}\label{3}
U(\mathbf{k}):=\exp({k}^a\,\hat{x}_a)U(\bm{0}).
\end{equation}
These coordinates would play the role of the conjugate momenta of
$\hat{x}_a$'s. The coordinates and the momenta in their operator forms
(denoted by hat) would satisfy the following relations
\begin{align}\label{4}
[\hat{k}^a,\hat{k}^b]&=0,\\
\label{5} [\hat{x}_a,\hat{k}^b]&=\hat{x}_a{}^b,
\end{align}
with ${x}_a{}^b$'s as scalar functions of $\hat{k}^a$'s,
having the property
\begin{equation}\label{6}
{x}_a{}^b(\mathbf{k}=\mathbf{0})=\delta_a^b.
\end{equation}
Accordingly, the operator forms of $\hat{x}_a$'s in the
$k$-basis are
\begin{equation}\label{7}
\hat{x}_a\to {x}_a{}^b\, \frac{\partial~}{\partial k^b}.
\end{equation}
The explicit forms of the $\hat{x}_a{}^b$ scalars
for the SU(2) group will be given later \cite{kfs}.
There are also the right-invariant vector fields
whose basis $\hat{x}_a^{\mathrm{R}}$ satisfy
\begin{align}\label{8}
[\hat{x}_a^{\mathrm{R}},\hat{x}_b^{\mathrm{R}}]&=
-f^c{}_{a\,b}\,\hat{x}_c^{\mathrm{R}},\\ \label{9}
[\hat{x}_a^{\mathrm{R}},\hat{x}_b]&=0.
\end{align}
Using these, one defines the vector field $\hat{J}_a$ through
\begin{equation}\label{10}
\hat{J}_a:=\hat{x}_a-\hat{x}_a^{\mathrm{R}}.
\end{equation}
Using (9) and the definitions of the left- and right-actions,
it is found that $J_a$'s are the generators of similarity transformation (adjoint action):
$$
\exp(\alpha^a \hat{J}_a)  U(\mathbf{k})= U(-\bm{\alpha}) U(\mathbf{k})
U(\bm{\alpha}).
$$
By all these above the following commutation relations hold
\begin{align}\label{11}
[\hat{J}_a,\hat{J}_b]&=f^c{}_{a\,b}\,\hat{J}_c,\\
\label{12} [\hat{J}_a,\hat{x}_b]&=f^c{}_{a\,b}\,\hat{x}_c,\\
\label{13} [\hat{k}^c,\hat{J}_a]&=f^c{}_{a\,b}\,\hat{k}^b.
\end{align}
According to the above commutation relations, in the case of the group SU(2)
($f^c{}_{a\,b}\!=\!\epsilon^c{}_{a\,b}$), $J_a$'s simply satisfy the commutation
relation of the rotation generators, and thus represent the components of the angular
momentum.

To construct the phase space, all that is needed is to transform
the commutation relations to Poisson brackets. This can be done
through the correspondence $[.\,,.]/(\ir\hbar)\to\{.\,,.\}$.
However, one should also take care of the dimension of the
quantities, and their reality (\textit{i.e.} Hermitian in operator form). To do so, let us define the
following quantities
\begin{align}\label{14}
p^a&:=(\hbar/\ell)\,\hat{k}^a,\\
\label{15} x_a&:=\ir\,\ell\,\hat{x}_a,\\
\label{16} x_a{}^b(\mathbf{p})&:=
\hat{x}_a{}^b[(\ell/\hbar)\,\mathbf{p}],\\
\label{17} J_a&:=\ir\,\hbar\,\hat{J}_a,
\end{align}
where $\ell$ is a constant of dimension length. One then arrives
at the following Poisson brackets
\begin{align}\label{18}
\{p^a,p^b\}&=0,\\
\label{19} \{x_a,p^b\}&=x_a{}^b,\\
\label{20} \{x_a,x_b\}&=\lambda\,f^c{}_{a\,b}\,x_c,\\
\label{21} \{J_a,x_b\}&=f^c{}_{a\,b}\,x_c,\\
\label{22} \{p^c,J_a\}&=f^c{}_{a\,b}\,p^b,\\
\label{23} \{J_a,J_b\}&=f^c{}_{a\,b}\,J_c,
\end{align}
where the dimension of $\lambda$ is that of inverse momentum:
\begin{equation}\label{24}
\lambda:=\frac{\ell}{\hbar}.
\end{equation}
One notes that $x_a$'s and $p^b$'s are independent variables, and
other variables can be expressed in terms of these. So that among
the Poisson brackets (\ref{18}) to (\ref{23}), only
(\ref{18}), (\ref{19}), and (\ref{20}) are
independent. All others can be derived from these.

Using (\ref{6}) it is seen that in the limit $\lambda\to 0$
(corresponding to $\ell\to 0$), the ordinary Poisson brackets are
retrieved.

\subsection{SU(2) setup}

For the special case of group SU(2),
the independent Poisson brackets
(\ref{18}), (\ref{19}), and (\ref{20}) are in fact the
Poisson structure of a rigid rotator, in which the angular
momentum and the rotation vector have been replaced by
$\mathbf{x}$ and $\mathbf{p}$, respectively, that is, the roles of
position and momenta have been interchanged.
It would be convenient to use the Euler parameters,
defined through
\begin{equation}\label{25}
\exp(\phi\,T_3)\,\exp(\theta\,T_2)\,\exp(\psi\,T_3):= \exp(k^a\,T_a),
\end{equation}
where $T_a$'s are the generators of SU(2) satisfying the
commutation relation
\begin{equation}\label{26}
[T_a,T_b]=\epsilon^c{}_{a\,b}\,T_c.
\end{equation}
It should be emphasized that in this setting the Euler angular parameters,
in opposite to their ordinary definition, are parametrizing the momentum space,
and their canonical conjugate components $X_\phi$, $X_\theta$, and $X_\psi$
play the role of the parametrization of the (real) configuration space.
One also mention that in this setup, $\phi$, $\theta$, and $\psi$
are dimensionless, while $X_\phi$, $X_\theta$,
and $X_\psi$ have the dimension of action. One advantage of using the Euler parameters over the
ordinary rotational generator $\bm{\theta}\cdot \mathbf{L}$ is that the Poisson brackets of $\phi$, $\theta$, $\psi$, $X_\phi$, $X_\theta$, and $X_\psi$ are the standard
canonical ones \cite{goldstein}, namely the only nonzero Poisson brackets are
\begin{align}\label{27}
\{X_\phi,\phi\}&=1,\\
\label{28} \{X_\theta,\theta\}&=1,\\
\label{29} \{X_\psi,\psi\}&=1.
\end{align}
Using this set of parameters one can find the basis for the generators of left action, whose
non-operator forms as spatial coordinates are found to be \cite{kfs}
\begin{align}\label{30}
x_1=\;&\lambda\,\left[-\frac{\cos\psi}{\sin\theta}\,X_\phi+\sin\psi\,X_\theta+
\frac{\cos\psi\,\cos\theta}{\sin\theta}\,X_\psi\right],\\
\label{31}
x_2=\;&\lambda\,\left[\frac{\sin\psi}{\sin\theta}\,X_\phi+\cos\psi\,X_\theta-
\frac{\sin\psi\,\cos\theta}{\sin\theta}\,X_\psi\right],\\
\label{32} x_3=\;&\lambda\,X_\psi.
\end{align}
Similarly the angular momentum components are found as follow \cite{kfs}
\begin{align}
\label{33}
J_1=\;&\frac{\cos\phi\,\cos\theta-\cos\psi}{\sin\theta}\,X_\phi+
(\sin\phi+\sin\psi)\,X_\theta \cr &+
\frac{-\cos\phi+\cos\psi\,\cos\theta}{\sin\theta}\,X_\psi,\\
\label{34}
J_2=\;&\frac{\sin\phi\,\cos\theta+\sin\psi}{\sin\theta}\,X_\phi
+(-\cos\phi+\cos\psi)\,X_\theta\cr &+
\frac{-\sin\phi-\sin\psi\,\cos\theta}{\sin\theta}\,X_\psi,\\
\label{35} J_3=\;&-X_\phi+X_\psi.
\end{align}
It is mentioned that the above components have the
correct dimension (\textit{i.e.} length for $x_a$'s and action for $J_a$'s).
One also has \cite{goldstein, kfs}
\begin{equation}\label{36}
\cos\frac{ k}{2}=\cos\frac{ \lambda\,p}{2}=\cos\frac{\theta}{2}\,
\cos\frac{\phi+\psi}{2},
\end{equation}
where $k:=\sqrt{\mathbf{k}\cdot\mathbf{k}}$ and
$p:=\sqrt{\mathbf{p}\cdot\mathbf{p}}$.

In the case with motion under a central force, the Poisson
brackets of Hamiltonian $H$ with $J_a$'s vanish. A Hamiltonian
which is a function of only $(\mathbf{p}\cdot\mathbf{p})$ and
$(\mathbf{x}\cdot\mathbf{x})$ is clearly so.
For such a system, $H$,
$\mathbf{J}\cdot\mathbf{J}$ and one of the components of
$\mathbf{J}$ (say $J_3$) are involutive constants of motion, hence
any SU(2)-invariant classical system is integrable.
As $\mathbf{J}$ is a constant vector, one can
choose the axes so that the third axis is parallel to this vector:
\begin{equation}\label{37}
J_1=J_2=0,
\end{equation}
by which, after subtracting $J_1\cos\phi+J_2\sin\phi$
and $J_1\cos\psi-J_2\sin\psi$, one has:
$$
\frac{(1+\cos\theta)}{\sin\theta}(1-\cos(\phi+\psi))(X_\phi-X_\psi)=0.
$$
Assuming $J_3=-X_\phi+X_\psi\neq0$, leads to
\begin{equation}\label{38}
\phi+\psi=0~\Rightarrow~ X_\phi+X_\psi=0.
\end{equation}
Defining the new variables
\begin{align}\label{39}
\chi&:=\frac{\psi-\phi}{2},\nonumber\\
J&:=-X_\phi+X_\psi,
\end{align}
for which $\{J,\chi\}=1$. So only $(J,\chi)$ and $(X_\theta, \theta)$ are left as the
canonically conjugate variables. Applying (\ref{38}), one arrives at
\begin{align}\label{40}
x_1&=\lambda\,\left(\frac{J}{2}\,\frac{1+\cos\theta}{\sin\theta}\,\cos\chi+
X_\theta\,\sin\chi\right),\nonumber\\
x_2&=\lambda\,\left(-\frac{J}{2}\,\frac{1+\cos\theta}{\sin\theta}\,\sin\chi+
X_\theta\,\cos\chi\right),\nonumber\\
x_3&=\lambda\,\left(\frac{J}{2}\right),\nonumber\\
\mathbf{x}\cdot\mathbf{x}&=\lambda^2\,\left[X_\theta^2+
\frac{J^2}{4}\,\left(1+\cot^2\frac{\theta}{2}\right)\right],\nonumber\\
\cos\frac{{k}}{2}&=\cos\frac{\theta}{2}.
\end{align}
It is seen that the motion is not in the plane $x_3=0$, but in a
plane parallel to that, as $x_3$ does not vanish but is a
constant.

By defining the polar coordinates ($\rho, \alpha$) in the $(x_1, x_2)$ plane:
\begin{align}\label{41}
x_1&=:\rho\,\cos\alpha,\nonumber\\
x_2&=:\rho\,\sin\alpha,
\end{align}
and $\mathbf{x}\cdot\mathbf{x}=x_1^2+x_2^2+x_3^2 $, one has
\begin{align}\label{42}
\mathbf{x}\cdot\mathbf{x}&=\rho^2+\frac{\lambda^2\,J^2}{4},\nonumber\\
\rho^2&=\lambda^2\,(X_\theta^2+J^2\,u^2),\nonumber\\
\alpha&=-\chi+\tan^{-1}\frac{X_\theta}{J\,u},
\end{align}
where
\begin{equation}\label{43}
u:=\frac{1}{2}\,\cot\frac{\theta}{2}.
\end{equation}
One then has
\begin{equation}\label{44}
\mathbf{x}\cdot\mathbf{x}=\lambda^2\,\left[X_\theta^2+
J^2\,\left(\frac{1}{4}+u^2\right)\right].
\end{equation}
In the line similar to the motion on ordinary space,
one can proceed to derive the equation for the path, in terms
of $\rho$ and $\alpha$. The first-order path equation
is found to be \cite{kfs}:
\begin{equation}\label{45}
\frac{1}{u^2}=\lambda^2\,J^2\,\left[\frac{1}{\rho^2}+\frac{1}{\rho^4}
\left(\frac{\d\rho}{\d\alpha}\right)^2\,\right].
\end{equation}
We mention that, by (\ref{40}) and (\ref{43}), $u$ can be expressed
as the kinetic energy. In the case of our interest, the $k$-space as angle variable
in (\ref{25}) is a compact one, and so
the momenta is taken to be bounded, that is $0\leq p \leq 2\pi /\lambda$.
Following \cite{0612013,fakE1,fakE2,kfs}, the kinetic term is taken to be
\begin{equation}\label{46}
K=\frac{4}{\lambda^2\,m}\,\left(1-\cos\frac{\lambda\,p}{2}\right)
\end{equation}
for a particle of mass $m$. By this choice, the kinetic term is a
SU(2)-invariant (rotationally-invariant), and hence a class-function
of SU(2), which is also monotonically increasing in the above mentioned
interval $0\leq p \leq 2\pi /\lambda$, just as in the case we have in
ordinary classical mechanics, with $p^2/(2\,m)$. Also we mention
in the limit $\lambda\to 0$, the expression (\ref{46})
coincides with the ordinary kinetic term, as it should.
One can express this kinetic term in terms of $u$:
\begin{equation}\label{47}
K=\frac{4}{\lambda^2\,m}\,\left(1-\frac{2\,u}{\sqrt{1+4\,u^2}}\right).
\end{equation}
Using above one easily find
\begin{equation}\label{48}
\frac{1}{u^2}=4\,\left[\left(1-\frac{\lambda^2\,m}{4}\,K\right)^{-2}-1\right].
\end{equation}
Assuming the usual form for the Hamiltonian as
\begin{equation}\label{49}
H=K+V(r),~~~~~ r:=\sqrt{\mathbf{x}\cdot\mathbf{x}}
\end{equation}
with $V(r)$ as the potential term, the path equation (\ref{45}) comes to the form
\begin{equation}\label{50}
\frac{1}{\rho^2}+\frac{1}{\rho^4}\left(\frac{\d\rho}{\d\alpha}\right)^2=
\frac{4}{\lambda^2\,J^2}\,\left\{\left[1-\frac{\lambda^2\,m}{4}\,\big(E-V(r)\big)\right]^{-2}
-1\right\},
\end{equation}
in which $r=\sqrt{\rho^2+x_3^2}=\sqrt{\rho^2+\lambda^2 J^2/4}$, by first of (\ref{42}).
It is easy to check that in the limit $\lambda\to 0$, the above
equation is reduced to the path equation on ordinary space. For the case of free particle,
it is easy to see that the solution of (\ref{50}) is of the form
$ \rho \cos (\alpha - \alpha_{0})=c$, representing a straight line \cite{kfs}.

\section{Circular and perturbed-circular orbits}
In this section the circular orbits under a central force are considered. In particular,
the existence condition for circular orbits, as well as the condition for stable
and closed slightly perturbed circular orbits are obtained.
Similar to the motion on ordinary space, it is useful to define the new variable:
\begin{equation}\label{51}
w:= \frac{1}{\rho}
\end{equation}
by which the path equation (\ref{50}) takes the form
\begin{equation}\label{52}
w'^2 + w ^2 =\frac{4}{\lambda ^2 J^2} \left\lbrace \left[ 1- \frac{\lambda ^2 m }{4} \big(E-V(r)\big) \right] ^{-2} -1 \right\rbrace
\end{equation}
with $w'=\d w /\d\alpha$, and
\begin{equation}\label{53}
r= \frac{1}{w} \sqrt{1+ \frac{\lambda ^2 J^2 w^2}{4}}.
\end{equation}
Differentiating once again with respect to $\alpha$ from (\ref{52}) we find
\begin{equation}\label{54}
 w''+w =\frac{-m}{J^2} \left[ 1- \frac{\lambda ^2 m}{4} \big( E-V(r) \big) \right] ^{-3} \frac{\d V(r)}{\d w}
\end{equation}

\subsection{Existence condition of circular orbits}
The circular orbit is given by the conditions
\begin{equation}\label{55}
\begin{array}{rl}
& w' \, =0, \\
& w'' \equiv 0.
\end{array}
\end{equation}
By using (\ref{52}) the first of above gives
\begin{equation}\label{56}
{w_0 }^2=  \frac{4}{\lambda ^2 J^2}\left\lbrace \left[ 1- \frac{\lambda ^2 m }{4}
\big(E-V(r_0)\big) \right] ^{-2} -1 \right\rbrace
\end{equation}
with $r_0$ given by (\ref{53}) at $w_0$. Above can
be converted to
\begin{equation}\label{57}
1+ \frac{\lambda ^2 J^2 w_0^2}{4} =\left[ 1- \frac{\lambda ^2 m }{4}
\big(E-V(r_0)\big) \right] ^{-2}.
\end{equation}
By using (\ref{54}) the second condition of (\ref{55}) leads to
\begin{equation}\label{58}
w_0 =\frac{-m}{J^2} \left[ 1- \frac{\lambda ^2 m}{4} \big( E-V(r_0) \big) \right] ^{-3}
\frac{\d V(r)}{\d w} \bigg|_{r_0}
\end{equation}
By combination (\ref{57}) and (\ref{58}), one finds the following as the condition
for existence of the circular orbit:
\begin{equation}\label{59}
w_0 = \frac{-m}{J^2} \left( 1+ \frac{\lambda ^2 J^2 w_0^2}{4}\right) ^{\frac{3}{2}}
\frac{\d V(r)}{\d w}\bigg|_{r_0}.
\end{equation}
The above relation is interpreted as an equation for $w_0$, whose positive solutions
($w_0 >0$) determine the radii of the allowed circular orbits. On ordinary space
the above condition takes the form \cite{goldstein}
\begin{equation}\label{60}
w_0 = \frac{-m}{J^2} \frac{\d V(r)}{\d  w}\bigg| _{\rho_0}~~~~\mathrm{(ordinary~ space)},
\end{equation}
which is simply retrieved as the limit $\lambda \rightarrow 0$ of (\ref{59}).
Using (\ref{53}),
\begin{equation}\label{61}
\frac{\d r}{\d w}=\frac{-1}{w^2}\left(1+\frac{\lambda ^2 J^2 w^2}{4}\right) ^{-1/2}
\end{equation}
by which one can replace for the derivative $\d V(r)/\d w$ appearing in (\ref{59}).
The above condition for the power law potentials of the form
\begin{equation}\label{62}
V(r)=\pm\, g\, r^n, ~~~~~~g>0,~~n>0
\end{equation}
leads to
\begin{equation}\label{63}
w_0 =\pm \frac{m}{J^2} n\, g\, w_0^{-n-1}\left( 1+\frac{\lambda ^2 J^2 w_0^2}{4}\right)^{(n+1)/2}
\end{equation}
or in a more compact form
\begin{equation}\label{64}
\frac{J^2}{m \rho _0}=\pm n\, g\, r_0^{n+1}
\end{equation}
showing that for the minus sign, for which the force is repulsive, the
circular orbit is not allowed. It is instructive to compare the above with
the condition on ordinary space:
\begin{equation}\label{65}
\frac{J^2}{m \rho _0}=  n\, g\, \rho_0^{n+1}~~~~\mathrm{(ordinary~ space)}.
\end{equation}
For the power law potentials of the form
\begin{equation}\label{66}
V(r)= \pm\, \frac{g}{r^n}, ~~~~~~~g>0,~~n>0
\end{equation}
the condition is found to be
\begin{equation}\label{67}
w_0= \mp \frac{m}{J^2} n\,g\, w_0^{n-1}
\left( 1+\frac{\lambda ^2 J^2 w_0^2}{4}\right)^{-n/2+1/2}
\end{equation}
or
\begin{equation}\label{68}
\frac{J^2}{m \rho _0}=\mp \frac{n\,g}{r_0^{n-1}}
\end{equation}
Again for the repulsive force (positive sign in (\ref{66})) the circular orbit is
not possible. In this case the condition on ordinary space is
\begin{equation}\label{69}
\frac{J^2}{m \rho _0}=\frac{n\,g}{\rho_0^{n-1}}~~~~\mathrm{(ordinary~ space)}.
\end{equation}

\subsection{Stable and closed perturbed circular orbits}
It is a matter of interest to look for the condition by which the circular orbits
stay bounded after a slight perturbation. To obtain the condition, we assume
a slight change in the parameter $w$ as
\begin{equation}\label{70}
w(\alpha)= w_0+w_1(\alpha)
\end{equation}
for which
\begin{equation}\label{71}
| w_1(\alpha)| \ll w_0.
\end{equation}
The purpose is to find the conditions under which the above
remains true not only as an initial condition, but also for all $\alpha$.
By inserting (\ref{70}) in (\ref{54}), we find the following as the
equation that $w_1(\alpha)$ should satisfy:
\begin{equation}\label{72}
w_1''(\alpha)+w_0+w_1(\alpha)=-\frac{m}{J^2}
\left\{ \left[1-\frac{\lambda ^2 m}{4}\big( E-V(r)\big) \right] ^{-3} \frac{\d V(r)}{\d w}\right\}
\bigg|_{r_0+r_1}.
\end{equation}
By expanding the right hand side of above in the first order of $w_1$, and using
(\ref{57}) and (\ref{59}), one finds the following for $w_1$
\begin{equation}\label{73}
w_1''(\alpha)+ \beta^2\, w_1(\alpha)=0,
\end{equation}
in which
\begin{equation}\label{74}
\beta ^2 :=1 -w_0\frac{\d}{\d w} \ln\left( -\frac{\d V(r)}{\d w}\right) \bigg|_{r_0}
-\frac{3 \lambda ^2 J^2}{4 \,r_0^2}.
\end{equation}
The condition for having a stable circular orbit is that
(\ref{73}) does not develop growing solutions in $\alpha$,
which is guaranteed if $\beta^2>0$. On ordinary space $\beta^2$ is
the same as above but in the limit $\lambda\to 0$ \cite{goldstein}.
The interesting difference between this model and the one on ordinary space comes
from the last term in above, by which, due to its negative sign, the cases with $\beta^2>0$
on ordinary space here may give a negative one.

For the attractive power law of the form
\begin{equation}\label{75}
V(r)=g\, r^n,
\end{equation}
for which the circular orbit is allowed, we find
\begin{equation}\label{76}
\beta ^2 = 1 + \frac{1+n}{r_0^2 w_0^2} ,
\end{equation}
which is always positive. The expression on ordinary space is gained in the limit
$\lambda\to 0$ (\textit{i.e.} $r_0^2 w_0^2\to 1$). For the attractive power law of the form
\begin{equation}\label{77}
V(r)=- \frac{g}{r^n}
\end{equation}
we then find
\begin{equation}\label{78}
\beta ^2 = 1 + \frac{1-n}{r_0^2 w_0^2} ,
\end{equation}
by which the $\beta^2>0$ condition, using (\ref{68}),  comes to the form
\begin{equation}\label{79}
n < 1+ r_0^2 w_0^2=2+\lambda^2 J^2 w_0^2/4.
\end{equation}
This result is less constrained than the condition on the ordinary
(\textit{i.e.} $n<2$). The interesting observation by (\ref{78})
is for the case $n=1$, the Kepler potential. In this case $\beta=1$ and
is independent of the initial condition. So, having (\ref{76}) and (\ref{78}),
one should expect that the only power law force for which not only the perturbed
circular orbits but also \textit{all} of the bound-states' orbits are closed could be the
Kepler one. Later we will see that this is indeed the case.

To have a closed perturbed circular orbit, $\beta$ should be a rational number \cite{goldstein}.
So whenever $\beta$ by (\ref{74}) would be a rational number, the perturbed circular orbit is
closed. On the ordinary space, for power law potentials the condition on $n$ for having closed orbits
is free from the initial conditions, and only depends on $n$ \cite{goldstein}.
Here, except for the case with $n=1$ in
(\ref{78}), due to existence of $r_0^2 w_0^2$ factor in (\ref{76}) and (\ref{78}),
both the value of $n$ and the initial conditions, by which $w_0$ is fixed, are important.

One can ask about the condition by which \textit{all} the perturbed circular orbits are closed.
Interpreting (\ref{74}) as a differential equation that potential $V(r)$ should satisfy for a fixed
rational $\beta_\mathrm{rat.}$, one would find the form of the potentials for which, no matter what
the initial conditions are, all of the perturbed circular orbits are closed \cite{goldstein}.
To do so one should write the condition (\ref{74}) in the way that the constant of motion
$J$ would not appear in it. In particular, the parameter $w$, which is related to 3-dimensional
radial variable $r$ by (\ref{53}), should be replaced by $r$. This can be done,
and the result comes to the form (see Appendix A)
\begin{equation}\label{80}
\beta_\mathrm{rat.}^2 =3 - \frac{1}{2}\lambda^2\, \mathcal{J}(r)
+ r\left(1-\frac{1}{4} \lambda^2\, \mathcal{J}(r)\right) \frac{\d}{\d\, r} \ln\left( -\frac{\d V(r)}{\d\, r}\right),
\end{equation}
in which
\begin{equation}\label{81}
\mathcal{J}(r):=\frac{m\,r}{8} \frac{\d V(r)}{\d\, r} \left(\sqrt{64+m^2\lambda^4 r^2
\left(\frac{\d V(r)}{\d\, r}\right)^2} - m\,\lambda^2 r \frac{\d V(r)}{\d\, r}\right).
\end{equation}

It is easy to see that the Kepler potential $V(r)=-g/r$ satisfies (\ref{80}) with
$\beta_\mathrm{rat.}=1$. In fact, by direct insertion, as expected by (\ref{78}),
it can be checked that, except the Kepler potential, the other kinds of attractive power
law potentials (\ref{75}) and (\ref{77}) do not satisfy the above equation. This observation
is in deep contrast with the case on ordinary space in which it can be shown that all the perturbed circular
orbits of the power law potentials in the form $V(r)\propto 1 / r^{2-\beta_\mathrm{rat.}^2}$
are closed \cite{goldstein}.

For a given $\beta_\mathrm{rat.}$, one still may look for the non-power law solutions of (\ref{80}).
The potentials by the numerical solutions of (\ref{80}) can be easily obtained and subsequently would be used in
the path equation (\ref{54}). In Figs.~1~\&~2 two examples of such potentials and their perturbed
circular orbit solutions are presented ($\beta_\mathrm{rat.}= 3/2 ~\&~ 5/2$).

\begin{figure}[t]
\includegraphics[scale=0.7]{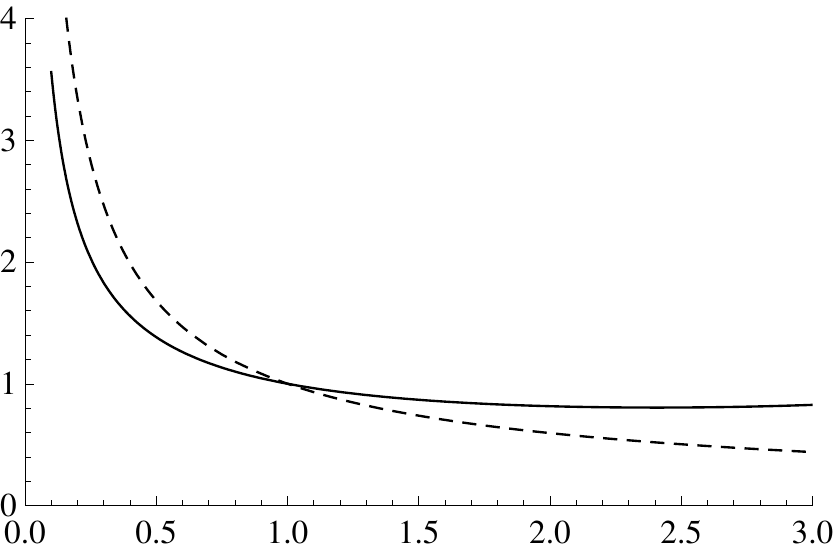}\hskip 1cm
\includegraphics[scale=0.5]{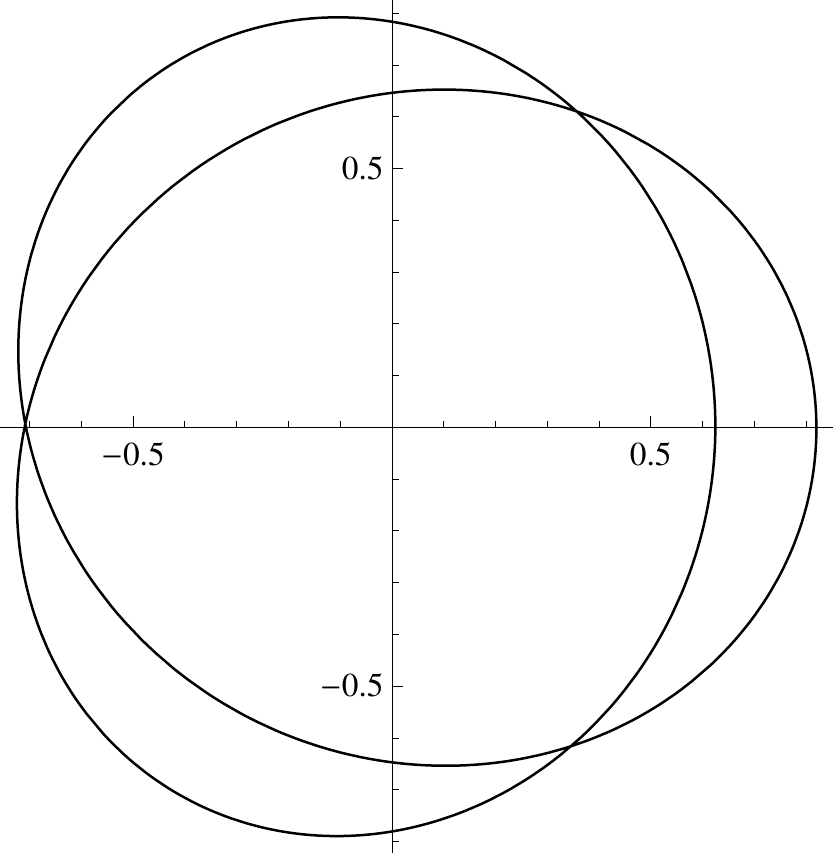}
\caption{{\small Left: The plot of $\d V(r)/\d r$ for $\beta_\mathrm{rat.}=3/2$, $m=2$, $\lambda=0.7$,
$\d V(r)/\d r|_{r=1}=1$; dashed-line:  $1/r^{3-\beta^2}=1/r^{0.75}$.
Right: The perturbed circular path is closed; $m=2$, $J=1$, $\beta=3/2$, $r(0)=1/1.6$, $r'(0)=0$. $r_\mathrm{circ.}\simeq1/1.4$ }}
\end{figure}

\begin{figure}[t]
\includegraphics[scale=0.7]{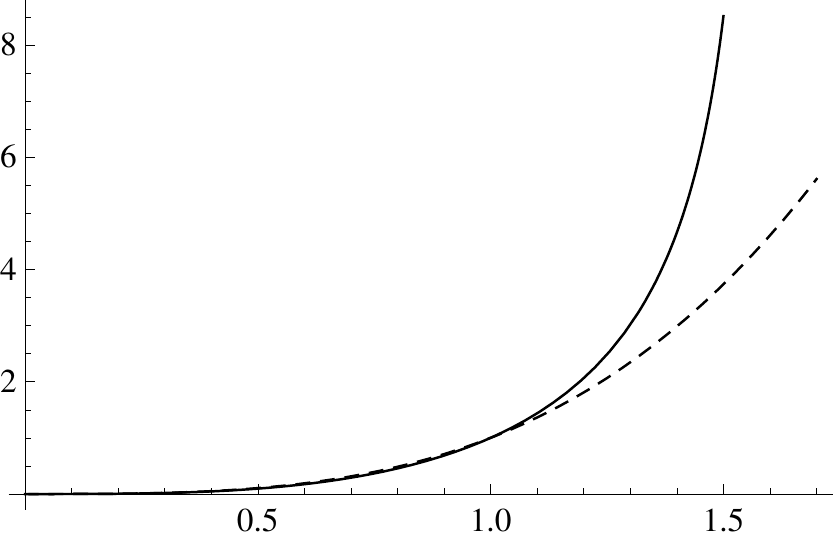}
\hskip 1cm
\includegraphics[scale=0.5]{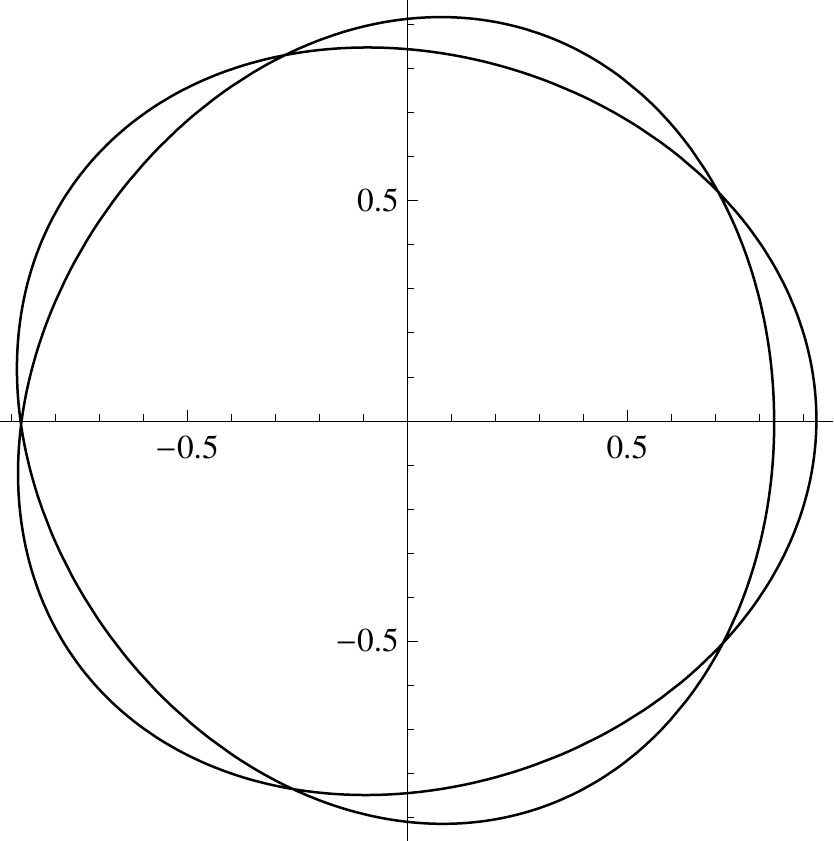}
\caption{{\small Left: The plot of $\d V(r)/\d r$ for $\beta_\mathrm{rat.}=5/2$, $m=2$, $\lambda=0.4$,
$\d V(r)/\d r|_{r=1}=1$; dashed-line: $1/r^{3-\beta^2}=r^{3.25}$.
Right: The perturbed circular path is closed; $m=2$, $J=1$, $\beta=5/2$, $r(0)=1/1.25$, $r'(0)=0$. $r_\mathrm{circ.}\simeq1/1.13$ }}
\end{figure}

One may go further and ask about the condition by which not only all the perturbed
circular orbits, but also all the orbits of bound-state solutions, no matter how far from circular
orbits, are closed. On ordinary space the so-called Bertrand theorem states that there are only two
potentials with the mentioned property, the Kepler potential ($-g/r$) and the harmonic oscillator one
($g\, r^2$) \cite{goldstein}. By the above considerations the latter one is excluded here (see also Fig.~3),
but the Kepler one is still a candidate for such a kind of potentials on SU(2) space. In the next section it is
shown explicitly that this is indeed the case. So on SU(2) space, among the power law potentials, it is the Kepler
problem that all the orbits of bound-state solutions are closed. Other potentials with the mentioned property, if any,
should be looked for among the non-power law solutions of (\ref{80}).

\begin{figure}[t]\begin{center}
\includegraphics[scale=0.8]{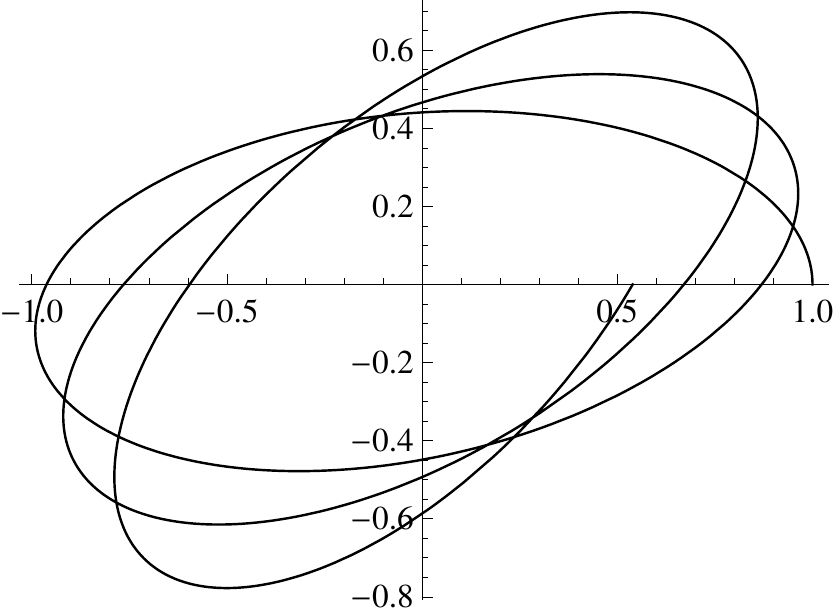}
\caption{{\small The path for harmonic oscillator potential $V(r)=g\,r^2$, with $m=2$, $J=1$, $g=1$, $\lambda=0.5$, $r(0)=1$, $r'(0)=0$.
The plot is for three revolutions and the path is not closed.  }}
\end{center}
\end{figure}

\section{Kepler potential: exact solution}
Here the solution to the equation (\ref{52}) for the Kepler potential is presented. The potential
in terms of the parameter $w$ comes to the form
\begin{equation}\label{82}
V(r)= -\frac{g}{r}= - g \, w  \left(1+\frac{\lambda^2 J^2 w^2}{4}\right)^{-1/2}.
\end{equation}
In the following, without lose of generality, we set $g=1$.
By above (\ref{52}) can be written as
\begin{equation}\label{83}
1+a^2 (w'^2 + w ^2) = \left[\eta - b \, w  \left(1+a^2 w^2 \right)^{-1/2} \right] ^{-2}.
\end{equation}
in which
\begin{equation}\label{84}
a:=\frac{\lambda J}{2},~~~b:=\frac{\lambda ^2 m }{4}g ,~~~
\eta:=1-\frac{\lambda ^2 m }{4}\,E.
\end{equation}
Using the new variable $v$ defined by
\begin{equation}\label{85}
a\,w(\alpha) =: \sinh v(\alpha),~~~ a \, w'(\alpha) = v'\,\cosh v(\alpha),
\end{equation}
(\ref{83}) comes to the form
\begin{equation}\label{86}
1+ v'^2 = \frac{1}{(\eta \cosh v - \gamma \sinh v)^2}
\end{equation}
in which $\gamma:=b/a$. So the dependence of the new variable
$v$ on the polar angle $\alpha$ can be obtained by
\begin{equation}\label{87}
\int \frac{\eta \cosh v - \gamma \sinh v}{\sqrt{1-(\eta \cosh v - \gamma \sinh v)^2}}\d v
= \pm(\alpha - \alpha_0),
\end{equation}
with $\alpha_0$ appearing as the constant of integration. The above integral
can be evaluated in two regimes: $\eta > \gamma$ and $\eta < \gamma$.
For the case with $\eta > \gamma$, introducing
$v_0:=\tanh^{-1}(\gamma/\eta)$ and $\mu:=1/\sqrt{\eta^2-\gamma^2}$, one has
\begin{equation}\label{88}
\eta \cosh v - \gamma \sinh v = \cosh(v-v_0)/\mu,
\end{equation}
by which the above integral takes the form
\begin{equation}\label{89}
\int \frac{\cosh(v-v_0)}{\sqrt{\mu^2-\cosh^2(v-v_0)}}\d v= \pm(\alpha - \alpha_0).
\end{equation}
The above integral is known \cite{gradsht}, leading to
\begin{equation}\label{90}
\sin^{-1} \frac{\sinh(v-v_0)}{\sqrt{\mu^2-1}}=\pm(\alpha - \alpha_0),
\end{equation}
or
\begin{equation}\label{91}
\frac{\sinh(v-v_0)}{\sqrt{\mu^2-1}}=\pm\sin(\alpha - \alpha_0).
\end{equation}
The last expression indicates that, through the intermediate variables $v$ and $w$, the
polar coordinate $\rho$ depends on $\sin(\alpha-\alpha_0)$. As the consequence,
the path after every revolution around the force center repeats itself, and hence is closed.

For the other case $\eta < \gamma$, again by introducing
$\tilde{v}_0:=\tanh^{-1} \eta/\gamma$ and
$\tilde{\mu}:=1/\sqrt{\gamma^2-\eta^2}$, one has
\begin{equation}\label{92}
\eta \cosh v - \gamma \sinh v =
\sinh(\tilde{v}_0-v)/\tilde{\mu},
\end{equation}
by which (\ref{87}) takes the form
\begin{equation}\label{93}
\int \frac{\sinh(\tilde{v}_0-v)}{\sqrt{\tilde{\mu}^2-\sinh^2(\tilde{v}_0-v)}}\d v
= \pm(\alpha - \alpha_0).
\end{equation}
The above integral is known \cite{gradsht}, leading to
\begin{equation}\label{94}
-\sin^{-1} \frac{\cosh(v-\tilde{v}_0)}{\sqrt{\tilde{\mu}^2+1}}=\pm(\alpha - \alpha_0),
\end{equation}
or
\begin{equation}\label{95}
\frac{\cosh(v-\tilde{v}_0)}{\sqrt{\tilde{\mu}^2+1}}=\mp\sin(\alpha - \alpha_0).
\end{equation}
The last expression indicates that the path is periodic and closed.
As a demonstration and to compare the path in the present case with
the one on ordinary space the orbits with equivalent parameters are plotted in Fig.~4.

\begin{figure}[t]\begin{center}
\includegraphics[scale=0.7]{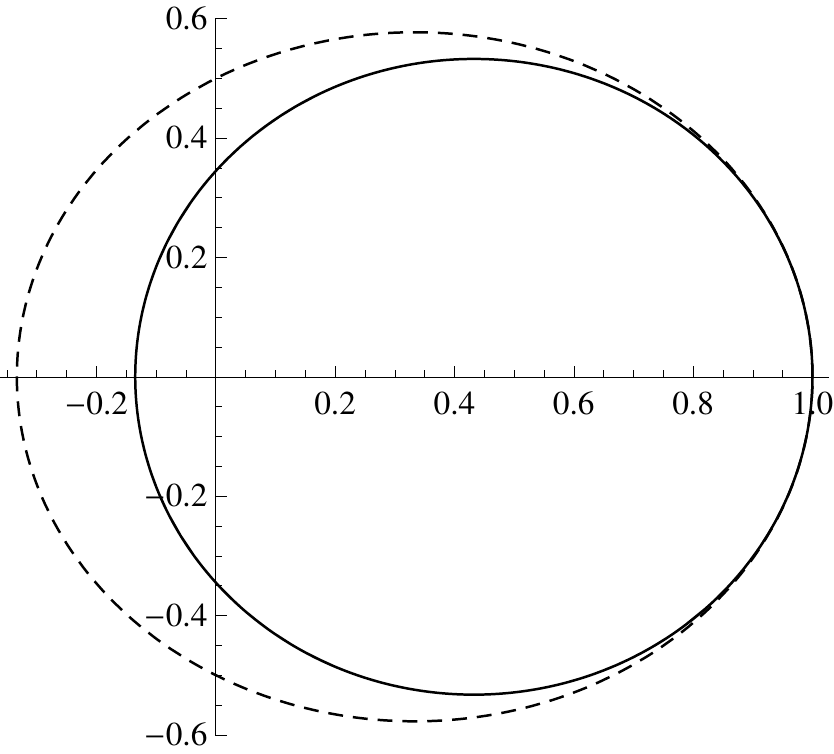}
\caption{{\small The sample plots of paths under Kepler potential; solid-line: SU(2) space; dashed-line: ordinary space. For both paths: $m=2$, $J=1$, $g=1$, $r(0)=1$, $r'(0)=0$.
For SU(2) path: $\lambda=2.1$.}}
\end{center}\end{figure}

It would be instructive to find the turning radii, in which the radial velocity vanishes
($w'=0$); the analogy of apsides of the elliptical path on ordinary space.
By (\ref{83}) one has the equation for the apsides
\begin{equation}\label{96}
1+a^2 w_\mathrm{aps.}^2 = \left[\eta - b\,  w_\mathrm{aps.} \left(1+a^2 w_\mathrm{aps.}^2 \right)^{-1/2} \right] ^{-2}
\end{equation}
or
\begin{equation}\label{97}
(1+b\,w_\mathrm{aps.})^2 =\eta^2 (1+a^2 w_\mathrm{aps.}^2 ),
\end{equation}
leading to
\begin{equation}\label{98}
w_\mathrm{aps.}=\frac{b\pm|\eta|\sqrt{a^2+b^2-a^2\eta^2}}{a^2\eta^2-b^2}.
\end{equation}
Restoring the original values the solutions come to the form
\begin{equation}\label{99}
w_\mathrm{aps.}=\frac{ 16\, m\,g \pm 2\, |4-m \lambda^2 E|\,
\sqrt{4\,m^2g^2 +m J^2 E (8 -m\lambda^2 E)}}{J^2 (4-m \lambda^2 E)^2 - 4\, m^2\lambda^2g^2}.
\end{equation}
The condition to have the real number roots is
\begin{equation}\label{100}
4\,m\,g^2 +  J^2 E\, (8 - m\lambda^2 E)\geq 0,
\end{equation}
by which, for fixed $J$, we have
\begin{equation}\label{101}
\frac{4}{m\lambda^2} \left(1- \sqrt{1+\frac{m^2 g^2\lambda^2}{4\,J^2}} \right)  \leq
E \leq  \frac{4}{m\lambda^2} \left(1+ \sqrt{1+\frac{m^2g^2 \lambda^2}{4\,J^2}} \right).
\end{equation}
The condition $-mg^2/(2J^2)\leq E$ on ordinary space can be retrieved
as the limit $\lambda\to 0$ of above. In the present case $E$ is bounded from both
below and above.

In the allowed region for $E$ only positive roots can be accepted as turning points. In the case
where both roots are positive the problem is in fact a bound-state one, and the roots
represent the apoapsis (smaller root) and the periapsis (larger root) of the path (remind $w=1/\rho$). In the cases where only one acceptable root exists, the problem
in fact is an unbounded motion, with the root representing the least distance to the center of force.

The circular path, if exists, is obtained by the condition $w_\mathrm{apoa.}=w_\mathrm{peri.}$,
which comes from (\ref{100}) by the equal-sign. Note the condition $4-m \lambda^2 E=0$ leads to
the unaccepted solution $w_\mathrm{apoa.}=w_\mathrm{peri.}<0$. On ordinary space the
condition for circular path is $E=-mg^2/(2J^2)$, which is obtained by $\lambda\to 0$ in (\ref{100}).

On the other extreme limit, one may look for the condition by which, for fixed $J$, the bound-state
with the farthest path from circle is obtained. On the ordinary
space this comes by the condition $E\to 0^-$, by which the apoapsis goes to $\infty$ ($w_\mathrm{apoa.}\to 0$) \cite{goldstein}.
In the present case,
setting $w'=0$ and $w\to 0$ in (\ref{83}), again we find $E\to 0^-$. Inserting $E\to 0^-$ in (\ref{99}), we find
\begin{equation}
w_\mathrm{peri./apoa.}\to\frac{ 4\, mg \pm 4 \, mg }{4\, J^2 - m^2\lambda^2g^2},
\end{equation}
which indicates bound-state paths with extremely large apoapsis are possible only for
$2 J > m\lambda g$. Otherwise, the apoapsis takes place in a finite value.

\appendix

\section{Derivation of (\ref{80}) and (\ref{81})}
The purpose is to replace $J$'s and $w$'s in (\ref{74}) by 3-dimensional
$r$. The staring point is (\ref{74})
\begin{equation}\label{}
\beta ^2 :=1 -w_0\frac{\d}{\d w} \ln\left( -\frac{\d V(r)}{\d w}\right) \bigg|_{r_0}
-\frac{3 \lambda ^2 J^2}{4 \,r_0^2}.
\end{equation}
First of all, using (\ref{53}) and (\ref{61}), the above can be brought to (\ref{80})
\begin{equation}\label{}
\beta ^2 =3 -\frac{\lambda ^2}{2}\frac{ J^2}{r_0^2}+ r_0 \left(1-\frac{\lambda ^2}{4}
\frac{J^2}{r_0^2}\right) \frac{\d}{\d\, r} \ln\left( -\frac{\d V(r)}{\d\, r}\right) \bigg|_{r_0}.
\end{equation}
Combining the relations (\ref{53}) and (\ref{61}), together with the circular orbit condition
(\ref{59}), one finds the following
\begin{equation}
\frac{1}{m^2 r_0^2 (\d V(r)/\d r|_{r_0})^2}\left(\frac{J^2}{r_0^2}\right)^2 +
\frac{\lambda ^2}{4} \frac{J^2}{r_0^2} -1=0
\end{equation}
by which we find (\ref{81})
\begin{equation}\label{}
\frac{J^2}{r_0^2}=\frac{m\,r}{8}\left.  \frac{\d V(r)}{\d\, r} \left(\sqrt{64+m^2\lambda^4 r^2
\left(\frac{\d V(r)}{\d\, r}\right)^2} - m\,\lambda^2r\, \frac{\d V(r)}{\d\, r}\right)\right|_{r_0}.
\end{equation}

\vspace{0.5cm} \noindent\textbf{Acknowledgement}:  This work is
supported by the Research Council of the Alzahra University.


\end{document}